%Paper: hep-ph/9410403
%From: cliff@hep.Physics.McGill.CA (Cliff Burgess)
%Date: Fri, 28 Oct 94 21:11:19 +0200

\font\tenrm=cmr10

\font\elevenbf=cmbx10 scaled\magstep 1
\font\elevenrm=cmr10 scaled\magstep 1
\font\elevenit=cmti10 scaled\magstep 1

\font\ninerm=cmr9

\hsize=6.0truein
\vsize=8.5truein
\parindent=1.5pc
\baselineskip=10pt

\line{hep-ph/9410403 \hfil McGill-94/01}
\vskip 1in
%==================================================================
\centerline{\elevenbf A RENORMALIZATION-GROUP APPROACH TO}
\vglue 7pt
\centerline{\elevenbf FINITE-TEMPERATURE MASS CORRECTIONS}
%===================================================================
\vglue 1.0cm
\centerline{\elevenrm A.L. MARINI\footnote{\tenrm$^\dagger$}{\ninerm Talk
presented at the 16$^{th}$ Annual Montr\'eal-Rochester-Syracuse-Toronto  (MRST)
 Meeting ``What Next? Exploring the Future of High-Energy Physcics'', held at
McGill University, Montr\'eal, Canada, 11-13 May 1994. To appear in Proceedings
published by World Scientific.} \elevenrm and C.P. BURGESS}
\baselineskip=13pt
\centerline{\elevenit Physics Department, McGill University, 3600 University
Street}
\baselineskip=12pt
\centerline{\elevenit Montreal, Quebec H3A 2T8, Canada}
%    IF NO SECOND AUTHOR, COMMENT OUT FOLLOWING LINES ----------
%  \vglue 0.3cm
%   \centerline{\elevenrm and}
%   \vglue 0.3cm
%\   \centerline{\elevenrm A.L. MARINI}
%  \centerline{\elevenit Group, Company, Address, City, State ZIP/Zone,
%               Country}
%==================================================================
\vglue 0.8cm
\centerline{\tenrm ABSTRACT}
\vglue 0.3cm
  {\rightskip=3pc
 \leftskip=3pc
 \tenrm\baselineskip=12pt
 \noindent
We illustrate how the reorganization of perturbation theory
at finite temperature can be economically cast in terms
of the Wilson-Polchinski renormalization methods.
We take as an example the old saw of the induced thermal
mass of a hot scalar field with a quartic coupling, which we
compute to second order in the coupling constant.  We show
that the form of the result can be largely determined by
renormalization-group arguments without the
explicit evaluation of Feynman graphs.
\vglue 0.8cm }
%================================================================
\line{\elevenbf 1. Introduction \hfil}
\vglue 0.2cm
%\line{\elevenit 1.1. Typeset Scripts \hfil}
%\vglue 0.1cm
\baselineskip=14pt
\elevenrm
For many years it has been known that finite-temperature perturbation theory
can be plagued by severe infrared divergences.  For example, amplitudes in
zero-temperature field theory in (3+1) dimensions are generally only
logarithmically divergent; whereas at finite temperature, the amplitudes can
diverge like a power of the infrared cutoff.  The source of this behaviour can
be traced to the appearance of the Bose-Einstein distribution function, which
behaves like $T/k$ for small momenta $k$.  The infrared divergences can be so
severe that the correspondence between the loop expansion and the
coupling-constant expansion is lost.  In general an infinite number of Feynman
graphs may contribute to a given order in the coupling-constant expansion$^1$.
In order to restore  perturbative calculability, a reorganization of
perturbation theory is required$^2$.  This reorganization can be achieved by
using a renormalization-group approach based on Wilson's formulation of the
effective action$^3$.  The purpose of this
paper is to show that one can use a renormalization-group approach to determine
the contributions to the induced thermal mass from the low-energy effective
theory without the explicit evaluation of Feynman graphs. This method also
allows one to extract the nonanalytic dependence on the coupling constant
which, in some cases, is numerically the most important contribution.
%\vglue 0.4cm
%\line{\elevenit 1.2. Section Headings \hfil}
%\vglue 0.1cm
%Section headings are to be in upper and lower case letters, and typeset
%in boldface. Sub-headings are to be in upper and lower case letters but
%typeset in italics. For each section or sub-heading, allow a space of about
%0.6 cm before it and 0.4 cm after it.
\vglue 0.6cm
\line{\elevenbf 2. Description of the example \hfil}
\vglue 0.4cm
As an example, consider the calculation of the induced thermal mass in a
massless scalar field theory with a quartic coupling.  The Lagrangian for the
system is given by
$${\cal L} = -{1\over{2}} \partial_\mu \phi \partial^\mu \phi - {g^2 \over{4!}}
\phi^4 \eqno(1)$$
and we work with the $(-+++)$ metric.  We use the Imaginary-Time Formalism$^4$
to perform the the calculation at finite temperature, where the boson energies
are given by $p_0=2 \pi i n T$ where $n$ is an integer and $T$ is the
temperature of the system.  Thus, the thermal scalar field can be described by
a Euclidean-space theory with the boson energies given by $p_0=2 \pi n T$ with
$n$ an integer.  We define the induced thermal mass as the real part of the
pole of the Minkowski-space propagator at zero three-momentum.  The induced
thermal mass can be expressed in terms of the Euclidean-space self-energy as
$$m^2 = - Re~\Pi({\bf p}=0)~~. \eqno(2)$$

We can now integrate out the high-energy modes in the theory described by
Eq.(1).  To do this we must integrate over all modes with $P^2>{\Lambda}^2$
where $P$ represents the four-momentum and $\Lambda$ is the scale which divides
the high-energy theory from the low-energy theory.  If we choose the scale
$\Lambda<T$, then all four-momenta in the effective theory must satisfy the
condition $p^2_0 + {\bf p}^2<{\Lambda}^2$ in Euclidean space. This condition
implies that we must integrate over all modes with $n\ne 0$ and integrate over
the $n=0$ modes with ${\bf p}^2>{\Lambda}^2$.  After performing such an
integration, one obtains the following effective Lagrangian:
$${\cal L}_\Lambda = -{Z_\Lambda\over{2}} \partial_\mu \phi \partial^\mu \phi -
{Z_\Lambda\over{2}}{m^2_\Lambda} \phi^2  - {{Z^2_\Lambda}{g^2_\Lambda}
\over{4!}} \phi^4 + {\rm other~terms}~~. \eqno(3)$$
The ``other terms'' include all possible combinations of the field and
derivatives thereof.  To compute the induced thermal mass with the effective
Lagrangian, one uses the new Feynman rules of the effective theory and
integrates over momenta  ${\bf p}^2<{\Lambda}^2$.  The induced thermal mass
cannot depend on the the scale $\Lambda$; therefore, we can write
$${d \over{d \Lambda}}~Z_\Lambda\Pi = 0 \qquad Z_\Lambda= 1 + O(g^4)\eqno(4)$$
with $\Pi= {m^2_\Lambda}+\delta {m^2_\Lambda}$ where $\delta {m^2_\Lambda}$
represents high-order corrections to ${m^2_\Lambda}$ computed with the
effective theory.  In general, one must include the wave-function
renormalization; however, since we are considering a calculation of the thermal
mass up to $O(g^4)$, the leading corrections to $Z_\Lambda$ can be ignored.  We
will refer to Eq.(4) as the renormalization-group equation.
\vglue 0.6cm
\line{\elevenbf 3. Induced thermal mass at one-loop order \hfil}
\vglue 0.4cm
If we integrate out the high-energy modes $P^2>{\Lambda}^2$ (where $P$
represents the four-momentum) to one-loop order, then the effective Lagrangian
is given by
$${\cal L}_\Lambda = -{1\over{2}} \partial_\mu \phi \partial^\mu \phi -
{1\over{2}}{m^2_\Lambda} \phi^2  - {{g^2_\Lambda} \over{4!}} \phi^4 + {\rm
other~terms} \eqno(5)$$
with
$${m^2_\Lambda}={g^2T^2\over{24}}-{g^2T\Lambda\over{4\pi^2}}+O(g^4)\qquad
{g^2_\Lambda}=g^2+\beta ( \Lambda ,T ) g^4+O(g^6)~~, \eqno(6)$$
and where the terms omitted from Eq.(5) are $O(g^4)$ and higher.  At this point
it is useful to define the following dimensionless variables:
$$\mu = {{m_\Lambda}\over{T}}\qquad \lambda={\Lambda\over{T}}\qquad
M^2={m^2\over{T^2}}\qquad z={\lambda\over{\mu}}~~.\eqno(7)$$

We can express the dimensionless induced thermal mass $M$ in the following
manner;
$$M^2 = \mu^2 + \mu g^2 S_1(z) + O(g^4,z) \eqno(8)$$
where $S_1(z)$ is an arbitrary function of $z$ that will be determined by
applying the renormalization-group equation given by Eq.(4).  The first
contribution in Eq.(8) is the direct thermal mass in the effective theory.  The
second term represents the one-loop correction to the direct thermal mass
computed within the effective theory, as illustrated in Figure 1(a).  The third
term represents corrections from diagrams with two or more loops.  The sum of
these contributions gives the induced thermal mass of the scalar field, which
must be independent of $\Lambda$.  The structure of the one-loop correction can
be easily understood.  The factor of $\mu$ is due to the fact that the loop
integral is three dimensional and there is only one propagator.  After one
factors out the direct thermal mass from the loop integral and divides by
$T^2$, a factor of $\mu$ will remain multiplied by a dimensionless integral
whose upper limit of integration is a function of the ratio $z=\lambda/\mu$.
By applying Eq.(4) to $M^2$,
 one can obtain the form of the function $S_1(z)$ without explicit evaluation
of the Feynman graph.  Applying the renormalization-group equation yields
$$-{g^2\over{4\pi^2}}-{{g^4z}\over{8\pi^2\lambda}}S_1(z) + {g^2}{d \over{d
z}}S_1(z)+{{g^4z^2}\over{8\pi^2\lambda}}{d \over{d
z}}S_1(z)+O\Bigl({g^4\over{\mu}},z\Bigr) =0 \eqno(9)$$
in which we have used the following relationships:
$${d\mu\over{d\lambda}}=-{{g^2z}\over{8\pi^2\lambda}}+O(g^4,z)\qquad
{dz\over{d\lambda}}={1\over{\mu}}
\Bigl(1+{{g^2z^2}\over{8\pi^2\lambda}}+
O(g^4,z)\Bigr)~~ \eqno(10)$$
where the derivatives are taken with $g$ fixed.  The solution of Eq.(9) for
$S_1(z)$ is easily obtained and found to be
$$S_1(z)={z\over{4\pi^2}} + C +O\Bigl({1\over{z}}\Bigr)~~ \eqno(11)$$
for large $z$.  This result can be compared with the explicit one-loop
calculation, which yields
$$\delta
{\mu}^2={g^2\mu\over{4\pi^2}}\int_{0}^{z}
{x^2dx\over{(x^2+1)}}={g^2\mu\over{4\pi^2}}(z-\arctan(z))
\eqno(12)$$
from which one can see that for $z>1$
$$S_1(z)={z\over{4\pi^2}} -{1\over{8\pi}}+O\Bigl({1\over{z}}\Bigr) \eqno(13)$$
which agrees with Eq.(11) with $C=-1/8\pi$.  Substitution of Eq.(13) into
Eq.(8) gives the induced thermal mass of the scalar field.  The value for the
mass is found to be
$$m^2=m_0^2\Bigl(1-{3m_0\over{\pi T}}\Bigr)+O(g^4)\qquad
m_0^2={g^2T^2\over{24}}\eqno(14)$$
which is in agreement with the result obtained by Dolan and Jackiw$^5$.  Their
calculation is performed with the original massless scalar field, and an
infinite sum of infrared-divergent ``daisy'' graphs is required to obtain the
finite result given by Eq.(14).  In our example, the perturbation theory has
been reorganized by integrating out the high-frequency modes and keeping the
direct thermal mass in the unperturbed sector of the Lagrangian.  To compute
the induced thermal mass with the effective theory, we only need to consider
one infrared-finite graph to obtain the leading-order correction.  By applying
the renormalization-group equation, we are able to determine the low-energy
contribution for large $z$ (up to integration constants) without explicit
evalution of the Feynman graph.  We will now extend our analysis to two-loop
order.
\vglue 0.6cm
\line{\elevenbf 4. Induced thermal mass at two-loop order \hfil}
\vglue 0.4cm
To calculate the induced thermal mass to $O(g^4)$, one needs to obtain the
effective Lagrangian to this order.  This can be achieved by integrating out
the high-frequency modes $P^2>{\Lambda}^2$ to two-loop order.  The result is
given by
$${\cal L}_\Lambda = -{1\over{2}} \partial_\mu \phi \partial^\mu \phi -
{1\over{2}}{m^2_\Lambda} \phi^2  - {{g^2_\Lambda} \over{4!}} \phi^4
-{cg^4\over{\Lambda^2}}{\phi^6\over{6!}}+ {\rm other~terms}\eqno(15)$$
with
$${m^2_\Lambda}={g^2T^2\over{24}}-{g^2T\Lambda\over{4\pi^2}}+
C(\Lambda,T)g^4+D(\Lambda,T)g^4+O(g^6)\qquad {g^2_\Lambda}=g^2+\beta ( \Lambda
,T ) g^4+O(g^6)~~ \eqno(16)$$
where $c$ is a constant.  The function $C(\Lambda,T)$ represents terms which
diverge in the limit $\Lambda \rightarrow 0$ whereas the function
$D(\Lambda,T)$ is finite in the limit $\Lambda \rightarrow 0$. At $O(g^4)$, an
effective six-point interaction must be included in the effective theory in
order for the effective theory to agree with the original massless theory$^6$.
Other dimension-six terms are possible; however, these operators have
derivative interactions.  By considering the lowest-order solution to the
classical equation of motion of the field, one can show that the dimension-six
terms with derivative interactions contribute to $O(g^6)$.  Following the same
technique as is used in Section 3, we can express the dimensionless induced
thermal mass as
$$M^2 = \mu^2 + \mu {g^2_\Lambda}
S_1(z)+{g^4_\Lambda}S_2(z)+g^4E+O\Bigl({g^6\over{\mu}},z\Bigr) \eqno(17)$$
where $E$ is a constant independent of $\lambda$.  The third term in Eq.(17)
represents the two-loop contribution to the mass constructed from two
four-point vertices.  The fourth term is the two-loop correction constructed
from the six-point interaction, which yields a constant at $O(g^4)$.  The final
term represents the contributions from diagrams with three or more loops.

After substituting the expressions for $\mu$, ${g^2_\Lambda}$, and $S_1(z)$
into Eq.(17) and applying the renormalization-group equation Eq.(4), the
following differential equation is obtained:
$${d\over{d\lambda}}\Bigl(g^4C(\lambda)+
{{g^2{\mu}^2}\over{4\pi^2\lambda}}\Bigr)+{{g^4z}
\over{64\pi^3\lambda}}+{g^4\over{\mu}}{d\over{dz}}S_2(z)
+{{g^6z^2}\over{8\pi^2\mu\lambda}}{d\over{dz}}S_2(z)+
O\Bigl({g^6\over{\lambda^2}},z\Bigr)=0 \eqno(18)$$
where $C(\lambda)$ is the dimensionless function at $O(g^4)$ obtained by
dividing $C(\Lambda,T)$, in Eq.(16), by $T^2$.  It is easy to verify that the
solution to Eq.(18) for $S_2(z)$ is given by
$$S_2(z) = -{1\over{64\pi^3}}z + C_1\log(z) + F+O\Bigl({1\over{z}}\Bigr)
\eqno(19)$$
for large $z$ where
$$C_1 = -\lambda {d \over{d
\lambda}}\Bigl(C(\lambda)+{1\over{96\pi^2\lambda}}\Bigr)~~. \eqno(20)$$

As was already shown, we can substitute the expression for $S_2(z)$ into
Eq.(17) and obtain the induced thermal mass to $O(g^4)$.  The induced thermal
mass of the scalar field is found to be
$$m^2=m_0^2\Bigl(1-{3m_0\over{\pi
T}}\Bigr)-g^4T^2C_1\log\Bigl({m_0\over{T}}\Bigr)+g^4T^2G+O(g^5)\qquad
m_0^2={g^2T^2\over{24}}~~.\eqno(21)$$
The form of the result given in Eq.(21) agrees with that found by Parwani$^7$
who computed the mass to $O(g^4)$ by explicit evaluation of the Feynman graphs
using the Imaginary-Time Formalism.  To calculate the constant $C_1$, one needs
to obtain the function $C(\lambda)$ by integrating out the high-frequency modes
in the two-loop Feynman graphs displayed in Figure 1.  After integrating, one
finds a $1/\lambda$ divergence in the contribution from Figure 1(b) and a
$\log(\lambda)$ divergence in the contribution from Figure 1(c).  The
$1/\lambda$ divergence in $C(\lambda)$ is precisely cancelled by the
$1/96\pi^2\lambda$ term appearing in Eq.(20).  Thus the constant $C_1$ is given
by the coefficient of the logarithmic term appearing in $C(\lambda)$ multiplied
by $-1$.  The value obtained is $C_1=-1/96\pi^2$, which also agrees with the
coefficient calculated by Parwani.  To determine the constant $G$ requires
explicit evaluation of the Feynman graphs.  In the small-coupling limit,
however, the logarithmic term
will dominate the $g^4T^2G$ term$^8$.

\input epsf.tex      %  only do this for the first figure in file
\vskip 0.6 cm       %  adjust as necessary for desired vertical spacing
                     %  of figure
\centerline{\epsfxsize 3.7 truein \epsfbox {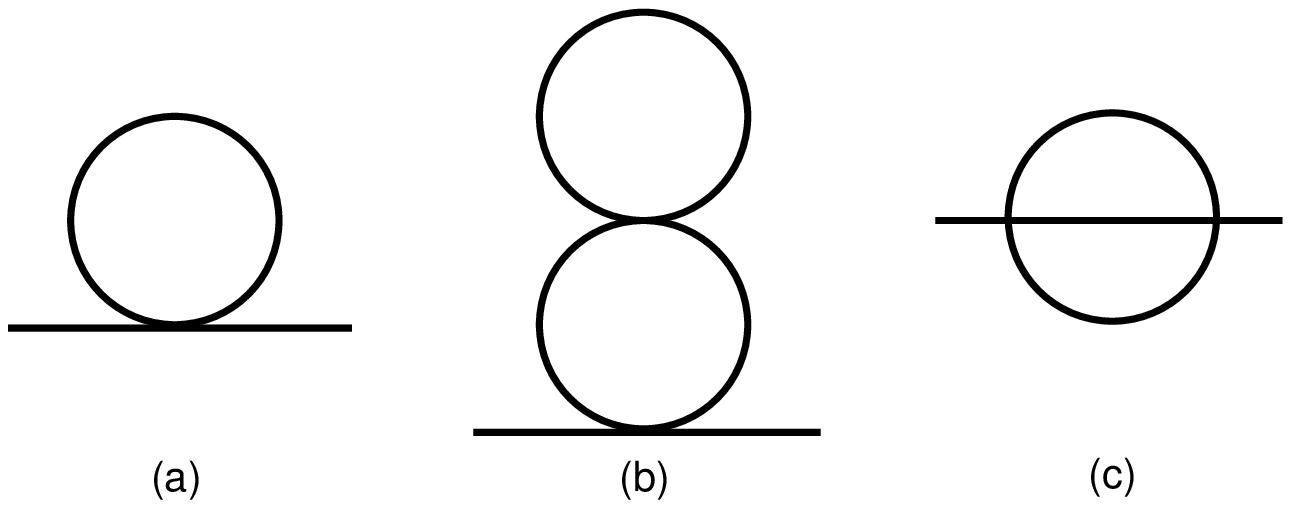}}
                     %  epsfxsize sets a scale (?), somehow works in
                     %   conjunction with the scale command within the
                     %   postscript file
                     %  also controls vertical placement of point
                     %   at which text resumes afterwards
                     %  Just leave above line as is.
       %  the rest is the figure caption.
       %  adjust spacings to suit taste:
\nobreak
\vskip 0.4 cm\nobreak      %   to adjust for where caption should appear
  %{ \narrower\narrower\narrower\smallskip
  %{Fig.~1:}\quad\quad{Figure caption}
\centerline {Fig. 1: Illustrations of one-loop and two-loop Feynman graphs.}
\smallskip
\vskip .1 cm
\vglue 0.6cm
\line{\elevenbf 5. Conclusions \hfil}
\vglue 0.4cm
We have shown how to use a renormalization-group approach to compute
finite-temperature mass corrections within the context of a scalar field
theory.  The method employed is very general and can be used to compute a wide
range of physical quantities in any theory.  This approach also provides a
framework to understand the resummation methods developed by Braaten and
Pisarski for hot QCD$^2$.  The resummed propagators and vertices given in their
resummed theory are analogous to the effective propagators and vertices that we
obtained by integrating out the high-frequency modes in the scalar field
theory.  It is hoped that the development of these methods will lead to a
better understanding of thermal field theories in general.

%================================================================
\vglue 0.6cm
\line{\elevenbf Acknowledgements \hfil}
\vglue 0.4cm
The authors would like to acknowledge helpful conversations with K. Dienes and
P. Labelle as well as funding by les fonds F.C.A.R. and the Walter C. Sumner
Memorial Foundation.
%==============================================================
\vglue 0.6cm
\line{\elevenbf References \hfil}
\vglue 0.4cm
%References in the bibliography should be referred to in the text by a
%superscript number without parentheses or brackets. All references
%should be organized to provide initials and last name of the author(s),
%title of publication (in italics), volume (in boldface), year of
%publication of paper in the journal/book and page numbers, e.g.,
\medskip
\item{1.} A. Linde, {\elevenit Rep. Prog. Phys.} {\elevenbf 42} (1979) 389, R.
Pisarski, {\elevenit Nucl. Phys.}
 {\elevenbf B309} (1988) 476.
\item{2.} E. Braaten and R. Pisarski, {\elevenit Nucl. Phys.}
 {\elevenbf B337} (1990) 569.
\item{3.} An interesting discussion concerning the renormalization-group
approach to quantum thermal field theories is given by  E. Braaten, in
{\elevenit Hot Summer Daze - BNL Summer Study on QCD at Nonzero Temperature and
Density}, eds. A. Gocksch and R. Pisarski (World Scientific, New Jersey, 1992).
\item{4.} T. Matsubara, {\elevenit Prog. Theor. Phys.}
 {\elevenbf 14} (1955) 351.
\item{5.} L. Dolan and R. Jackiw, {\elevenit Phys. Rev.}
 {\elevenbf D9} (1974) 3320.
\item{6.} J. Polchinski, {\elevenit Nucl. Phys.}
 {\elevenbf B231} (1984) 269.
\item{7.} R. Parwani, {\elevenit Phys. Rev.}
 {\elevenbf D45} (1992) 4695.
\item{8.} C.P. Burgess and A.L. Marini, {\elevenit Phys. Rev.}
 {\elevenbf D45} (1992) 17; A. Rebhan, {\elevenit Phys. Rev.}
 {\elevenbf D46} (1992) 482.
%\vglue 0.6cm
%\line{\elevenbf\noindent 7. Footnote\hfil}
%vglue 0.4cm
%Footnotes should be typeset in 9 point roman at the bottom
%of the page where it is cited.

\bye